\documentclass{aa}
\usepackage{txfonts}
\usepackage{graphicx}

\usepackage[normalem]{ulem}

\usepackage{natbib}
\bibpunct{(}{)}{;}{a}{}{,} 

\newcommand{\be}{\begin{equation}}
\newcommand{\ee}{\end{equation}}
\newcommand{\bea}{\begin{eqnarray}}
\newcommand{\eea}{\end{eqnarray}}
\newcommand{\p}{\partial}
\newcommand{\vphi}{{\cal V}^{(\phi)}}

\let\phi=\varphi
\let\rho=\varrho

\begin{document}

\headnote{Research Note}
\title{Mass estimate of the XTE~J1650-500 black hole from the Extended Orbital Resonance Model for high-frequency QPOs
}
\titlerunning{Mass estimate of the XTE~J1650-500}

\author{Petr Slan\'{y} \and Zden\v{e}k Stuchl\'{\i}k} 
\institute{Institute of Physics, Faculty of Philosophy and Science, Silesian
  University in Opava, Bezru\v{c}ovo n\'{a}m. 13, CZ-74601 Opava, Czech Republic}

\offprints{P. S., \email{petr.slany@fpf.slu.cz}}

\date{Received / Accepted}

\abstract
{XTE~J1650-500 is a Galactic black-hole binary system for which at least one high-frequency QPO at 250\,Hz was reported. Moreover there are indications that the system harbours a near-extreme Kerr black-hole with a spin $a_{\ast}\simeq 0.998$ and mass $M_{\rm BH}\lesssim 7.3\,M_{\sun}$. Recently it was discovered that the orbital 3-velocity of test-particle (geodesical) discs orbiting Kerr black holes with the spin $a_{\ast}>0.9953$, being analyzed in the locally non-rotating frames, reveals a hump near the marginally stable orbit. Further it was suggested that the hump could excite the epicyclic motion of particles near the ISCO with frequencies typical for high-frequency QPOs. Characteristic frequency of the hump-induced oscillations was defined as the maximal positive rate of change of the LNRF-related orbital velocity with the proper radial distance. If the characteristic ``humpy frequency'' and the radial epicyclic frequency are commensurable, strong resonant phenomena are expected.
}
{Application of the idea of hump-induced oscillations in accretion discs around near-extreme Kerr black holes to estimate the black-hole mass in XTE~J1650-500 binary system. 
} 
{For the Kerr black hole with spin $a_{\ast}\simeq 0.9982$ the characteristic ``humpy frequency'' and the radial epicyclic frequency at the orbit, where the positive rate of change of the LNRF-related orbital velocity with the proper radial distance is maximal, are in ratio 1:3. Identifying the radial epicyclic frequency with the observed 250\,Hz QPO, we arrive at the mass of the black hole. In this method the ratio of frequencies determines the spin (and vice versa), and concrete values of frequencies determine the black-hole mass.
}
{Mass of the Kerr black hole in XTE~J1650-500 binary system is estimated to be around $5.1\,M_{\sun}$.
} 
{}

\keywords{Accretion, accretion discs -- Black hole physics -- Relativity -- X-rays: individuals: XTE~J1650-500}

\maketitle


\section{Introduction}
The Galactic X-ray transient binary black-hole system XTE J1650-500 belongs among a few sources in which a near-extreme stellar-mass Kerr black hole is assumed (another one is, e.g.,  the microquasar GRS~1915+105). \citet{Mil-etal:2002:ASTRJ2:} observed the source with \textsl{XMM-Newton}/EPIC-pn device close to the peak of its outburst in September 2001, and found a broad, skewed Fe K$\alpha$ emission line in its spectrum, indicating a stellar-mass Kerr BH with near-maximal angular momentum (spin) close to the value $a_{\ast}\simeq 0.998$. In fact, the spin $a_{\ast}=0.998$ corresponds to the well-known ``Thorne limit'', which is the upper limit on spin of the ``Laor'' model \citep{Lao:1991:ASTRJ2:} used by Miller et al. to fit the line. The ``Thorne limit'' is a very serious consequence of theoretical considerations how photons, coming from the inner part of a geometrically thin, geodesical accretion disc around the Kerr black hole, influence a spin-up process of the hole induced by accreting matter on the hole. Depending on the photon emission law the limiting value slightly differs from the ``canonical value'' $a_{\ast}\simeq 0.998$, being 0.9978 for isotropic emission and 0.9982 for the electron-scattering emission law, see \citet{Tho:1974:ASTRJ2:} for more details. Indicia for such a high spin in XTE~J1650-500 was found also in three observations with \textsl{BeppoSAX} during the same outburst, which were performed just before and after the \textsl{XMM-Newton} observation, see \citep{Min-Fab-Mil:2004:MONNR:}. On the other hand \citep{Don-Ger:2006:MONNR:} suggests that if the absorption features in the optically thick hot corona are taken into account, the inferred relativistic smearing of the K$\alpha$ line can be substantially reduced to be compatible with the conception of a disc truncated at $\sim 10\,r_{\rm g}$ ($r_{\rm g}=GM_{\rm BH}/c^2$ is a gravitational radius), accomplished by the evaporating hot inner accretion flow, and thus giving no information on the black-hole spin. In this paper we analyze one consequence of the possibility that the black hole in XTE~J1650-500 is very rapidly rotating.

\citet{Hom-etal:2003:ASTRJ2:} reports detection of a high-frequency variability in X-ray flux from XTE~J1650-500, analyzing data from the \textsl{Rossi X-Ray Timing Explorer} obtained from September to November 2001. During transition of the source from the hard to the soft state, the high-frequency QPO around 250\,Hz (with maximum observed frequency of 270\,Hz) was detected. It is worth to say that \citet{Hom-etal:2003:ASTRJ2:} reports also broad high-frequency features around 50\,Hz, 109\,Hz, and 168\,Hz, among which the feature around 50\,Hz is remarkably stable. Similarly, \citet{Kal-etal:2003:ASTRJ2:} reports also broad features near 80\,Hz (but with a quite large error of $\sim 50\%$) and 25\,Hz, but no high-frequency QPO. It is not clear whether this 80\,Hz peak is the same as the 50\,Hz peak found by \citet{Hom-etal:2003:ASTRJ2:}. Now we can only say that the data revealed some broad high-frequency features with frequencies of tens Hz, and possibly one high-frequency QPO around 250\,Hz. 

Subsequently, \citet{Oro-etal:2004:ASTRJ2:} published their analysis of the R-band photometry of the binary system XTE~J1650-500, determining the optical mass-function, $f(M)=(2.73\pm 0.56)\,M_{\odot}$, and a~lower limit to the inclination of the system of $50^{\circ}$. The results give an~upper limit to the mass of the black hole in XTE~J1650-500 of $M_{\rm BH}\lesssim 7.3\,M_{\odot}$ \citep{Oro-etal:2004:ASTRJ2:}.

We summarize that the spectral and timing properties of the X-ray transient black-hole candidate XTE~J1650-500 suggest that this source hosts a stellar-mass (and possibly near-extreme) Kerr BH in its center. 

\section{Aschenbach effect in thin, geodesical discs}
\begin{figure}
\includegraphics[width=1 \hsize]{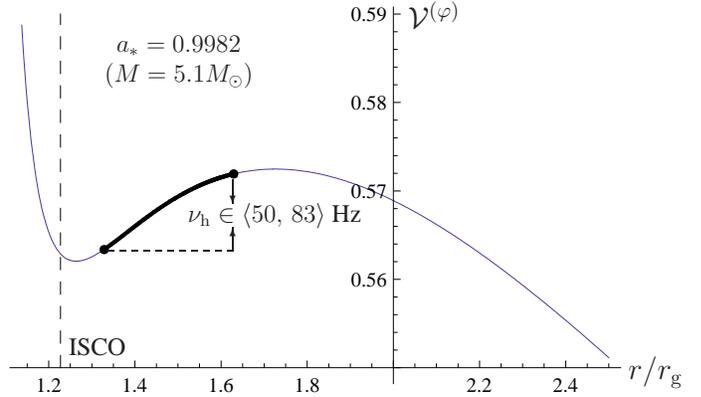}	
\caption{Orbital velocity profile along equatorial circular geodesics (in units of $c$) determined by the locally non-rotating frames in the case of a very rapidly rotating Kerr black hole with a dimensionless spin $a_{\ast}=0.9982$. The positive part of the radial gradient of $\vphi$ is located close to but above the innermost stable circular orbit (ISCO). The thick part corresponds to the region where the humpy frequency, determined for the $5.1 M_{\odot}$ black hole, reaches the values from the interval 50--83\,Hz, corresponding to observed broad high-frequency features in XTE~J1650-500.}
\label{f1}
\end{figure}

In the case of very rapidly rotating Kerr black holes (with the dimensionless spin $a_{\ast}=cJ/GM_{\rm BH}^{2}>0.9953$; $J$ is the black-hole angular momentum), a test particle orbital 3-velocity $\vphi$, defined by appropriate projections of particle's 4-velocity $U^{\mu}$ onto the LNRF-tetrad \citep{Bar-Pre-Teu:1972:ASTRJ2:},\footnote{In rotating spacetimes a rotation of the space, causing so-called ``dragging of inertial frames'', is superposed on the own orbital motion of matter in the disc, as viewed from infinity. Locally non-rotating frames (LNRF) are dragged along with the spacetime, thus the LNRF should reveal \emph{local} orbital properties of the disc in the clearest way, similarly as the static observers do in non-rotating spacetimes.}
reveals a non-monotonic profile in the equatorial plane \citep{Asch:2004:ASTRA:}. \citet{Stu-Sla-Tor-Abr:2005:PHYSR4:} shown that the analogous humpy behaviour of the $\vphi$ takes place also for non-geodesic motion of a test perfect fluid orbiting near-extreme Kerr black holes with spin $a_{\ast}>0.9998$ in marginally stable thick discs (tori), characterized by uniform distribution of the specific angular momentum, $\ell(r,\,\theta)=-U_{\phi}/U_{t}=\mbox{const}$.\footnote{Standard Boyer-Lindquist (B-L) coordinates $(t,\,r,\,\theta,\,\phi)$ are used.} In both, geometrically thin and thick accretion discs, the positive part of $\p\vphi/\p r$ is confined to the ergosphere around the black hole, however it is located above the innermost stable circular orbit (ISCO) in the equatorial plane ($\theta=\pi/2$); see Fig.~\ref{f1}. Therefore local processes in the inner part of accretion discs around near-extreme Kerr black holes could carry a signature of the orbital velocity hump. 

\citet{Asch:2004:ASTRA:} introduced a characteristic (critical) frequency of any process connected with the velocity hump by the maximum positive slope of the orbital velocity $\vphi$, being the function of the B-L radius $r$. Its coordinate-independend definition, using the maximal positive rate of change of $\vphi$ with the \textsl{proper} radial distance $\tilde{r}$, 
\be                                                                             \label{e1}
     \nu_{\rm crit}^{\tilde{r}}=\frac{\p\vphi}{\p\tilde{r}}|_{\rm max}, \quad
     \mathrm{d}\tilde{r}=\sqrt{g_{rr}}\mathrm{d} r,
\ee
was given by \citet{Stu-Sla-Tor:2004:RAGtime4and5:}. Relating this locally defined characteristic frequency $\nu_{\rm crit}^{\tilde{r}}$ to a static observer at infinity, we get so-called ``humpy frequency'' \citep{Stu-Sla-Tor:2007a:ASTRA:}
\be                                                                             \label{e2}
     \nu_{\rm h}=\sqrt{-(g_{tt}+2\omega g_{t\phi}+\omega^2 g_{\phi\phi})} 
     \nu_{\rm crit}^{\tilde{r}},
\ee
where $g_{\mu\nu}$ are metric coefficients of the Kerr geometry (in B-L coordinates), and $\omega=-g_{t\phi}/g_{\phi\phi}$ is the angular velocity of the LNRF \citep{Bar-Pre-Teu:1972:ASTRJ2:}. It should be stressed, however, that the frequency, characterizing the Aschenbach effect in the disc, need not be fixed and defined strictly as the \textsl{maximal} gradient of the orbital velocity in the proper radial distance, Eq.~(\ref{e1}). Rather we expect that the characteristic frequency can slightly change and be shifted from this maximal value. On the other hand, at the orbit where it is maximal, i.e. at the so-called ``humpy radius'', the Aschenbach effect is the strongest. Due to possible changes of the humpy frequency the Aschenbach effect in the disc should not be characterized by a~narrow frequency peak but rather by one relatively wide and not very strong QPO with centroid frequency of several tens Hz (for stellar mass Kerr black holes).

\citet{Asch:2004:ASTRA:} also gave a heuristic assumption on possible excitation of particle's epicyclic motion in the inner part of geodesical discs by the orbital velocity hump. The epicyclic motion is characterized by the radial and vertical epicyclic frequencies $\nu_{\rm r},\,\nu_{\theta}$ (for their explicit definition in the more general Kerr-Newman spacetime see \citet{Ali-Gal:1981:GENRG2:}). \citet{Stu-Sla-Tor:2007a:ASTRA:} shown that for $a_{\ast}\to 1$ the humpy frequency and the epicyclic frequencies, being evaluated at the ``humpy radius'', reach asymptotic, i.e. almost spin-independent, commensurable values, being in the ratios $\nu_{\theta}\!:\!\nu_{\rm r}\!:\!\nu_{\rm h}\simeq 11\!:\!3\!:\!2$. 
Moreover, in the region with positive slope of $\vphi$ the orbits of commensurability between vertical and radial epicyclic frequencies $r_{3:1}$ and (for $a_{\ast}\gtrsim 0.996$) $r_{4:1}$ are located, see \citep{Asch:2004:ASTRA:,Stu-Sla-Tor:2007a:ASTRA:}, supporting the idea of connection between the humpy profile of $\vphi$ and epicyclic motion of particles. In general, it is also possible to expect an excitation of oscillations in the disc by the orbital velocity hump; the concrete physical mechanism is, however, unknown.

\begin{figure}
\includegraphics[width=1 \hsize]{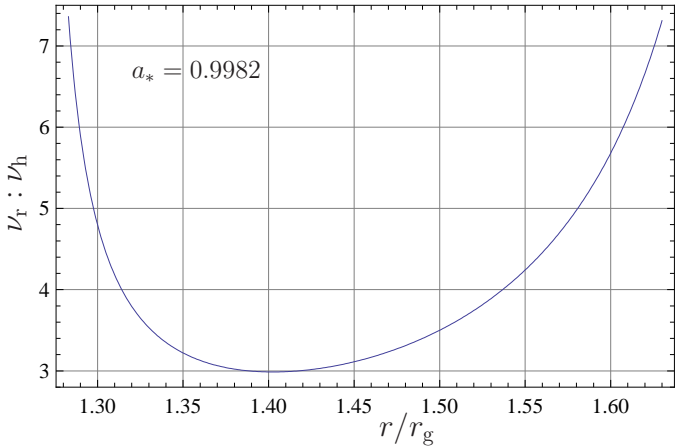}	
\caption{Ratio of the radial epicyclic frequency and the humpy frequency for equatorial geodesical orbits around the Kerr black hole with spin $a_{\ast}= 0.9982$.}
\label{f2}
\end{figure}

Model in which the resonance between the hump-induced and epicyclic motion is assumed, so-called Extended Orbital Resonance Model (ExORM), was applied to the X-ray variable Galactic black hole GRS~1915+105, in which at least five high-frequency QPOs with frequencies $\nu_{1}=27$\,Hz \citep{Bel-Men-San:2001:ASTRA:}, $\nu_{2}=41$\,Hz, $\nu_{3}=67$\,Hz \citep{Mor-Rem-Gre:1997:ASTRJ2:,Str:2001:ASTRJ2:}, $\nu_{4}=113$\,Hz, $\nu_{5}=167$\,Hz \citep{Rem:2004:AIPC:} were observed.
The model predicts near-extreme spin of the central black hole, $a_{\ast}=0.9998$, which is in agreement with results given by spectral continuum fits, $a_{\ast}>0.98$, presented by \citet{McCli-etal:2006:ASTRJ2:}, and the black-hole mass $M_{\rm BH}\sim 14.8\,M_{\sun}$, being
well inside the interval $(14.0\pm 4.4)\,M_{\odot}$ given by other observational methods \citep{Rem-McCli:2006:ARASTRA:}. The results are partly summarized in Fig.~\ref{f3} and fully described in our previous paper \citep{Stu-Sla-Tor:2007b:ASTRA:}. 

\section{Mass estimate of the XTE~J1650-500 black hole}
As mentioned above, the very broad Fe-line observed in XTE~J1650-500 is usually interpreted as a~signature of a~near-extreme Kerr black hole presented in the system. \footnote{Note that some other alternatives were proposed to explain such broadness, e.g. \citep{Don-Ger:2006:MONNR:}.} If the source really contains the black hole spinning close to the `Thorne limit' $a_{\ast}\simeq 0.998$, the effects of non-monotonic orbital velocity profile could take place there. 

For test-particle (geodesical) discs and the spin $a_{\ast}= 0.9982$, ratio of the radial epicyclic frequency and the characteristic ``humpy frequency'', both computed at the same radius, is presented in Fig.~\ref{f2}. Of special interest are the ratios given by small integers like 3:1, 4:1, and 5:1, which can play a role in resonances between hump-induced and epicyclic oscillations. We assume that the most probable is the forced resonance, however, other kinds of resonances are not completely excluded. From the resonant point of view the most relevant is the 3:1 ratio, which can be excited through the superharmonic resonance due to some non-linearity of the 3-rd order that could be assumed in the system \citep{Nay-Moo:1995:NonlinearOscillations:}. Moreover, strength of the resonance decreases with rising order of the non-linearity \citep{Lan-Lif:1976:Mechanics:}, being stronger for the 3:1 ratio than for the 4:1 or 5:1 ratio. Further it should be stressed that the 3:1 resonant radii are located at the orbits very close to the ``humpy radius'' $r_{\rm h}\simeq 1.417r_{\rm g}$, i.e., the orbit where the characteristic ``humpy frequency'' reaches its maximal value given by Eq.~(\ref{e2}), and where the Aschenbach effect is the strongest. Locations of resonant radii 3:1, 4:1, and 5:1 can be inferred from Fig.~\ref{f2} and are explicitly given in Table~\ref{t1}.

Now we identify the strong 250\,Hz QPO with the radial epicyclic frequency at given resonant radius and compute corresponding black-hole mass. At preferred ``humpy radius'', which almost coincides with the outer resonant orbit of the 3:1 resonance, the radial epicyclic frequency reaches the value of $1275\, (M/M_{\odot})^{-1}$\,Hz, which gives the mass of the Kerr black hole $M_{\rm BH}\simeq 5.1\,M_{\odot}$. Mass-estimates corresponding to other resonant radii are given in Table~\ref{t1}.
\begin{table}
	\centering
		\begin{tabular}{ccccc}\hline\hline
		$\nu_{\rm r}\!:\!\nu_{\rm h} $ & $r_{\rm in}/r_{\rm g}$ & $r_{\rm out}/r_{\rm g}$ & $M_{\rm in}/M_{\odot}$ & $M_{\rm out}/M_{\odot}$ \\ \hline
		3:1 & 1.388 & 1.417 & 4.7 & 5.1 \\
		4:1 & 1.314 & 1.537 & 3.3 & 6.6 \\
		5:1 & 1.298 & 1.581 & 3.0 & 7.0 \\	\hline\hline
		\end{tabular}
	\caption{Location of resonant orbits (inner, outer) corresponding to ratios $\nu_{\rm r}$:$\nu_{\rm h}=$\,3:1,\,4:1, and 5:1 for the Kerr black hole ($a_{\ast}= 0.9982$) schematically given in Fig.~\ref{f2}. Next two columns show mass-estimates for the black hole in XTE~J1650-500, if the radial epicyclic frequency at each orbit is identified with the observed 250\,Hz QPO.}	
	\label{t1}
\end{table}

\section{Conclusions}
Suggested application of the Extended Orbital Resonance Model to observed quasi-periodic variability in the X-ray transient XTE~J1650-500, like other strong-gravity orbital models of high-frequency QPOs, e.g. the `Relativistic Precession Model' of \citet{Ste-Vie-Mor:1999:ASTRJ2:} or the orbital resonance model of \citet{Klu-Abr:2001:ACTPB:}, is able to give the estimates for the mass and spin of the Kerr black hole, if more than one frequency or one of black-hole parameters are known. Of course, the employed ideas can be relevant only if the source harbours a near-extreme Kerr black hole, because only in this case the LNRF-related orbital velocity of the accretion disc reveals the non-monotonic behavior in its inner part close to the ISCO. 

\begin{figure}
\includegraphics[width=1 \hsize]{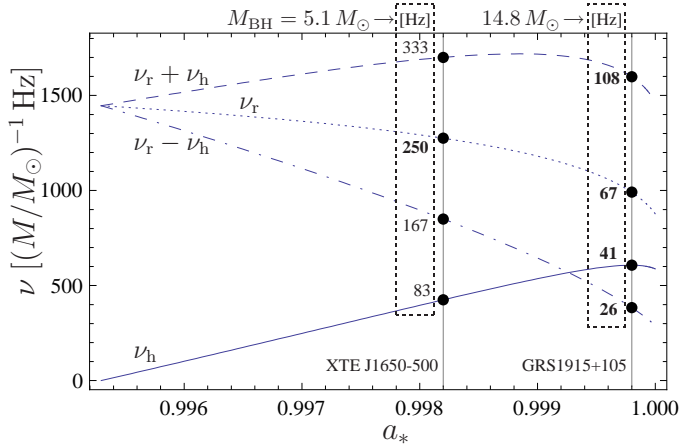}	
\caption{Spin dependence of the humpy and radial epicyclic frequency, $\nu_{\rm h}$ and $\nu_{\rm r}$, and their combinational frequencies $\nu_{\rm r}\pm\nu_{\rm h}$. All frequencies are calculated at the \emph{same} radius, where the rate of change of the orbital velocity in terms of the proper radial distance reaches the maximal positive value for a given $a_{\ast}$. 
The values of frequencies, corresponding to the black-hole mass and spin implied by suggested explanations of X-ray variability in XTE~J1650-500 and GRS~1915+105 binary black-hole systems, are given. The bold-faced values can be identified with the observed ones, referred as QPOs for given source.}
\label{f3}
\end{figure}

Following the spectral and timing properties we assume that the spin of the Kerr black hole in XTE~J1650-500 is equal to the `Thorne limit' $a_{\ast}= 0.9982$ for the electron-scattering photon emission law, and that the observed high-frequency QPO at 250\,Hz corresponds to the epicyclic motion in the orbital plane, excited by the orbital velocity hump. The first assumption about the spin relies on the analyses of the broad Fe K$\alpha$ line \citep{Mil-etal:2002:ASTRJ2:, Min-Fab-Mil:2004:MONNR:} which suggest the spin close to the `Thorne limit'. The second assumption about the nature of the 250\,Hz QPO is model-dependent. Further we expect the existence of resonances between the hump-induced and epicyclic motions. Under these assumptions we evaluated ratios of the radial epicyclic frequency and the characteristic humpy frequency in the region of positive radial gradient of $\vphi$, and found particular resonant radii. The most promising seems the resonance $\nu_{\rm r}:\nu_{\rm h}\sim 3:1$, since it takes place at the ``humpy radius'' where the positive rate of change of the orbital velocity with the proper radial distance is maximal and its strength is much higher in comparison to resonances with 4:1 or 5:1 frequency ratio \citep{Lan-Lif:1976:Mechanics:}. In the case of 3:1 ratio at the humpy radius the inferred mass of the black hole in XTE~J1650-500 is $5.1\,M_{\odot}$, which is well below the upper limit $7.3\,M_{\odot}$ given by \citet{Oro-etal:2004:ASTRJ2:}. The humpy frequency and corresponding combinational frequencies $\nu_{\rm r}\pm\nu_{\rm h}$, naturally existing in non-linear systems, are given in Fig.~\ref{f3}. Recall that some broad high-frequency features near frequencies 80\,Hz \citep{Kal-etal:2003:ASTRJ2:} and 168\,Hz \citep{Hom-etal:2003:ASTRJ2:} were reported, and could be identified with $\nu_{\rm h}$ and $\nu_{\rm r}-\nu_{\rm h}$, respectively.\footnote{Recently, at the 10th HEAD meeting held in Los Angeles, California from March 31st to April 3rd, 2008, N.~Shaposhnikov \& L. Titarchuk presented another estimate of the mass of the XTE~J1650-500 black hole obtained by Quasi Periodic Oscillation - Spectral Index scaling method, being $(3.8\pm 0.5)\,M_{\odot}$, which is also in good agreement with dynamical measurements and remarkably close to the expected neutron star maximum mass.} 

General character of the humpy frequency, characterizing the `Aschenbach effect' in accretion discs, enables to expect that the real frequency of any process connected with the orbital velocity hump differs from the maximal value for $\nu_{\rm h}$, and, in general, is not fixed to any particular value, depending, e.g., on accretion rate and internal processes in the inner part of the disc. For $(5.1\,M_{\odot},\ 0.9982)$ Kerr black hole the range of radii, for which the humpy frequency reaches values between 50\,Hz and 83\,Hz, is emphasized in Fig.~\ref{f1}. This frequency interval, containing the observed broad high-frequency features in XTE~J1650-500, covers almost the whole region with positive radial gradient of $\vphi$.

The main goal of this note is to show that the observed spectral and timing properties of the X-ray transient XTE~J1650-500 black hole could be interpreted in terms of the Extended Orbital Resonance Model, giving the possibility to estimate the mass of the black hole.
Nevertheless, the presented results should not be read as definite, rather they should be taken as another contribution to the discussion about the character of the black hole in XTE~J1650-500, supporting some previously published results and emphasizing possible existence of the new effect in accretion discs around near-extreme Kerr black holes discovered by Aschenbach four years ago.


\begin{acknowledgements}
The authors acknowledge useful discussions with Gabriel T\"{o}r\"{o}k. 
The work has been done as a part of the research project MSM~4781305903 financed by the Czech Government.
\end{acknowledgements}

\bibliographystyle{aa} 


\end{document}